%% file: main_arXiv.tex
\documentclass[twocolumn,prl,superscriptaddress]{revtex4-1}
\pdfoutput=1
\usepackage{amsmath,breqn,mathtools,amsfonts} 
\usepackage{hyperref}
\allowdisplaybreaks
\usepackage{graphicx}
\usepackage{tikz}
\usepackage{xcolor}
\usepackage{bm}
\usepackage{rotating}
\usepackage[export]{adjustbox}

\DeclarePairedDelimiterX{\ExpArg}[1]{[}{]}{#1}

\newcommand{\bv}[1]{\mathbf{#1}}
\newcommand{\vecc}[1]{\boldsymbol{#1}}

\makeatletter
\let\cat@comma@active\@empty
\makeatother
\begin{document}

\title{Out of equilibrium response and fluctuation-dissipation violations across scales  in flocking systems}
\author{Federica Ferretti}
\altaffiliation{Current address: NSF-Simons National Institute for Theory and Mathematics in Biology, 60611 Chicago, IL}
\affiliation{Department of Chemical Engineering, Massachusetts Institute of Technology, Cambridge, Massachusetts 02139, USA}
\author{Irene Giardina}
\email[Corresponding author: ] {irene.giardina@uniroma1.it}
\affiliation{Dipartimento di Fisica, Universit\a`{a} Sapienza, 00185 Rome, Italy}
\affiliation{Istituto Sistemi Complessi, Consiglio Nazionale delle Ricerche, UOS Sapienza, 00185 Rome, Italy}
\affiliation{INFN, Unit\a`{a} di Roma 1, 00185 Rome, Italy}
\author{Tomas Grigera}
\affiliation{Instituto de Física de Líquidos y Sistemas Biológicos (IFLySiB), Universidad Nacional de La
Plata and CONICET, La Plata, Argentina}
\affiliation{CCT CONICET La Plata, Consejo Nacional de Investigaciones Científicas y Técnicas, La Plata,
Argentina}
\affiliation{Departamento de Física, Facultad de Ciencias Exactas, Universidad Nacional de La Plata,
La Plata, Argentina}
\affiliation{Istituto Sistemi Complessi, Consiglio Nazionale delle Ricerche, UOS Sapienza, 00185 Rome, Italy}
\author{Giulia Pisegna}
\affiliation{Max Planck Institute for Dynamics and Self-Organization (MPI-DS), D-37077 Goettingen, Germany
}
\author{Mario Veca}
\email[Corresponding author: ]{ mario.veca@uniroma1.it}
\affiliation{Dipartimento di Fisica, Universit\a`{a} Sapienza, 00185 Rome, Italy}

\date{\today}

\begin{abstract}
Flocking systems are known to be strongly out of equilibrium. Energy input occurs at the individual level to ensure self-propulsion, 
and the individual motility in turn contributes to ordering, enhancing information propagation and strengthening collective motion. However, even beyond ordering, a crucial feature of natural aggregations is response. How, then, do off-equilibrium features affect the response of the system? In this work, we consider a minimal model of flocking and investigate response behavior under directional perturbations.
We show that equilibrium dynamical fluctuation-dissipation relations between response and correlations are violated, both at the local and at the global level. The amount of violation peaks at the ordering transition, exactly as for the entropy production rate.  Entropy is always produced locally and connected to the local fluctuation-dissipation violation via Harada-Sasa relationships. 
However, cooperative mechanisms close to the transition spread off-equilibrium effects to the whole system, producing an out of equilibrium response on the global scale. Our findings elucidate the role of activity and interactions in the cost repartition of collective behavior and explain what observed in experiments on natural living groups.

\end{abstract}


\maketitle

\noindent
{\bf Introduction.}
Collective motion in living systems has been observed on different scales and regimes, from swimming bacteria and cell colonies to insect swarms and bird flocks \cite{krause_book}. From the perspective of statistical physics, these phenomena represent paradigmatic instances of non-equilibrium emergent behavior and they have been intensively studied, both at theoretical and experimental level, in the context of active matter \cite{marchetti2013hydro,ramaswamy2010mechanics,gompper20202020}. Broadly speaking, a flock has the following distinctive features: individuals are `active', i.e. they are endowed with a self-propulsion mechanism transforming energy into motion, and they coordinate with each other via short range interaction rules.  Individual motility is responsible for the non equilibrium character of the problem. Interactions produce collective patterns on the large scale. The combination of these two ingredients leads to a non-trivial phenomenology comprising kinetic ordering transitions, novel classes of critical behavior and motility induced phase separation
\cite{vicsek1995novel,vicsek2012collective,chate2008collective,cates2015motility}.

 The impact of activity on order has been investigated at length. Seminal works \cite{vicsek1995novel,tonertu1995,tonertu1998} have indeed shown that self-propulsion favours ordering, by enhancing information propagation through the system and stabilizing the transition to collective motion.  On the other hand, much less is known on how non-equilibrium features affect the response of the system to perturbations. Response is a key trait for survival in living aggregations, and it therefore represents a quantity of primary biological relevance. Besides, even more than order itself, it provides a {\it bona fide} signature of collective behavior, by quantifying the ability to retain global coherence in presence of external signals or threats. In this work, we address systematically this question. We consider an archetypal model of flocking and compute analytically and numerically the response function. We show that 
equilibrium dynamical fluctuation-dissipation relations between response and correlations are violated on all timescales, and we quantify the corresponding deviation. The global amount of the non-equilibrium contribution depends on the noise and it peaks at the transition point. We explain how non-equilibrium effects distribute across scales: energy is injected at the individual level via self-propulsion and violation of detailed balance occurs on the scale of the interaction range; however, close to the transition  cooperative mechanisms arise in the dynamics and the non-equilibrium effects spread on all scales.

\vskip 0.1 cm

\noindent
{\bf The model and the perturbation protocol.}
We consider a continuous time generalization of the Vicsek model of flocking \cite{vicsek1995novel,chepizhko2021revisiting} where self-propelled particles regulate their own speed using a speed control potential, and interact with each other locally via mutual adjustment of their velocities \cite{bialek2014,cavagna2022marginal}. The model captures in minimal terms the imitative nature of behavioral coordination rules in aggregations of living organisms \cite{krause_book}, as well as the effective interactions present in many non-living active assemblies \cite{vicsek2012collective}. The equations of motion  are
\begin{equation}
    \frac{d \mathbf{v}_i}{dt} = - \frac{\partial \mathcal H}{\partial \mathbf{v}_i} + \boldsymbol{\xi}_i \  \ \ ,
  \ \    \frac{d \mathbf{r}_i}{dt} = \mathbf{v}_i
\label{eom}
\end{equation}
where $\boldsymbol{\xi}_i$ is a white noise with variance $ \langle \xi_{i\alpha}(t) \xi_{j\beta}(t') \rangle = 2 T \delta_{ij} \delta_{\alpha \beta}\delta(t-t')$, 
and the pseudo Hamiltonian ${ \mathcal H}$ reads
\begin{equation}
   \mathcal{H} = \frac{J}{2 } \sum_{ij} n_{ij}(\mathbf{v}_i -\mathbf{v}_j)^2 + \frac{g}{2} \sum_i (|\mathbf{v}_i| - v_0)^2 \ .
   \label{ham}
\end{equation}
 The first term in this expression describes mutual adjustment between velocity vectors, $J$ being the scale of the interaction and $n_{ij}(t)$ the adjacency matrix.  
  In the following, we will adopt a metric  interaction rule, thereby $n_{ij}(t)=1$ if particles $i$ and $j$ at time $t$ have mutual distance smaller than a fixed range $r_0=1$, and $n_{ij}(t)=0$ otherwise. The second term in Eq.~\eqref{ham} is the speed control, setting the individual speeds close to a cruising reference value $v_0$, with fluctuations regulated by the parameter $g$ \footnote{
 We could have chosen different forms for the speed potential (e.g. functions of ${\bv v}_i^2$ rather than $|\mathbf{v}_i|$). In the region around the ordering transition - which is the most interesting regime for our analysis - there is no much difference, so we stick to the simple Gaussian shape of Eq.~\ref{ham}. }.
 The fixed modulus Vicsek model  is recovered for $g \to \infty$.

 With decreasing noise strength, model (\ref{eom}) exhibits a kinetic ordering transition to a state of collective motion, where the mean group velocity ${\bv V} = (1/N) \sum_i {\bv v}_i$  and the global degree of alignment - the polarization $\vecc{\Phi}=(1/N) \sum_i \bv{v}_i/|\bv{v}_i|$  - are different from zero (see App. \ref{app:numerical}).
Our aim here is to investigate response behavior, and to understand how it relates to ordering and to the out-of-equilibrium features of the model. 
To do so, we consider a perturbation protocol where external fields are applied to the individual velocities, i.e.
 $\mathcal{H}  \to  \mathcal{H} -\sum_i {\bv h}_i \cdot {\bv v}_i$.
  The same protocol  has been recently used 
  in \cite{chate2008collective,kyriakopoulos2016leading,brambati2022signatures,loffredo2023collective}); however, no systematic investigation has been performed so far of fluctuation-dissipation relations.

\vskip 0.1 cm 

\noindent
{\bf Response functions and fluctuation-dissipation relations.}
The linear response functions 
can be defined in the usual way as $R^{\alpha\beta}_{i j}(t,s) =  \delta \langle v_i^\alpha(t)\rangle_h/\delta h_j^\beta(s) |_{h=0}$
 \cite{marconi2008fluctuation}. 
In equilibrium systems, such response functions are related to the velocity correlation functions $C^{\alpha\beta}_{ij}(t,s) = \langle v_i^\alpha(t)  v_j^\beta(s)   \rangle $  by the dynamical Fluctuation Dissipation Theorem (FDT) \cite{kubo1966fluctuation}. However, model (\ref{eom}) is intrinsically out of equilibrium due to the active nature of the particles, 
and we  expect violations of such relationships. 
In the stationary state, for the diagonal spacial components (the only ones that, for symmetry reasons, are non-zero in our case), we can then write (for $t>s)$
\begin{equation}
T R^{\alpha \alpha}_{i j}(t-s) =  -  \frac{d C^{\alpha\alpha}_{ij}(t-s)}{dt} - d^{\alpha \alpha}_{ij}(t-s) 
\label{rviolation}
\end{equation} 
where $d^{\alpha \alpha}_{ij}$  measures the deviation from equilibrium.

In general, computing either experimentally or numerically the full response tensor is a demanding task, as one would need to apply local fields and monitor reactions for all pairs of individuals. On the contrary, {\it global responses} are more easily accessible and statistically more reliable. Let us then consider the simpler setup where we apply a uniform field $h_i^\alpha (t) = h_\alpha(t)$ on the system and we measure how the global mean velocity ${\bv V}$   
 is affected by the perturbation. In this case,  
  the global linear response is given by $R_{\alpha \alpha}(t-s) = \delta \langle V_\alpha(t) \rangle / \delta h_\alpha(s)|_{h=0}= (1/N) \sum_{ij} R^{\alpha \alpha}_{i j}(t-s)$, and the analogue of Eq.~(\ref{rviolation}) is
\begin{equation}
T R_{\alpha \alpha}(t-s) =  -  \frac{d C_{\alpha\alpha}(t-s)}{dt} -  d_{\alpha \alpha}(t-s)  \ ,
\label{rviolationg}
\end{equation} 
with $C_{\alpha\alpha}(t-s)= N \langle V_\alpha(t) V_\alpha(s)\rangle$ and $d_{\alpha \alpha}(t-s)= (1/N)\sum_{ij}d_{ij}^{\alpha \alpha}(t-s)$.

Interestingly, for Langevin dynamics the response functions can be explicitly computed starting from the equations of motion. To this task, it is convenient to introduce the Onsager-Machlup action \cite{onsager1953fluctuations} $S[\{\bv{v}_i, \bv{r}_i\}]$, defined by 
\begin{align}
& P\left [\{\bv{v}_i(t),\bf{r}_i(t)\} \right ]  =   \frac{1}{\cal N} \exp{\{-S[\{\bv{v}_i(t),\bv{r}_i(t)\}]\}}   \label{OM}  \\
& \quad=  \biggl \langle \delta  \left ( {\bv v}_i(t)-{\bv v}_i(0)-\int_0^t dt' \ [- \partial{\cal H}/\partial{\bv v}_i +{\vecc \xi}_i (t')] 
 \right )\biggr \rangle \ .
  \nonumber
\end{align}
where  averages are performed over the  trajectory realizations. 
In terms of the action $S$, the response functions can be written as  $R^{\alpha\alpha}_{i j}(t,s) = - \langle v_i^\alpha(t)\ \delta S / \delta h_j^\alpha(s) \rangle |_{h=0}$.
Using standard techniques of stochastic calculus \cite{cugliandolo2011effective,marconi2008fluctuation,caprini2021generalized}
one then finds, for $t>s$,  (see App. \ref{app:OM} for details)
\begin{equation}
 \label{viola}
 d_{\alpha \alpha} = 
  \frac{gN v_0}{2} [ \langle V_\alpha(t) \Phi_\alpha(s) \rangle - \langle V_\alpha(s) \Phi_\alpha(t) \rangle ] \ .
\end{equation}
The deviation from equilibrium is now expressed in terms of the global correlation between velocity and polarization, computed in zero field.
Eq.~(\ref{viola}) therefore allows to precisely quantify the FDT violation and to compute the response (via Eq.~\eqref{rviolationg}) 
 without actually perturbing the system. We note that in the r.h.s of Eq.~(\ref{viola}) the speed control parameter $g$ explicitly appears. In fact, as we will show later on, the deviation only mildly depends on $g$ and a well defined asymptotic $g\to \infty$ limit is quickly reached as soon as $g\sim{\rm O}(1)$. Still, allowing the individual speed to fluctuate is crucial to derive Eq.~(\ref{viola}). In other terms, speed fluctuations are not indispensable to produce out-of-equilibrium effects, but they can be used very effectively to capture and quantify them
  (see App.\ref{app:equilibrium} for more discussion).

\vskip 0.1 cm

\begin{figure}[t]
\centering
 \includegraphics[width=0.47\textwidth]{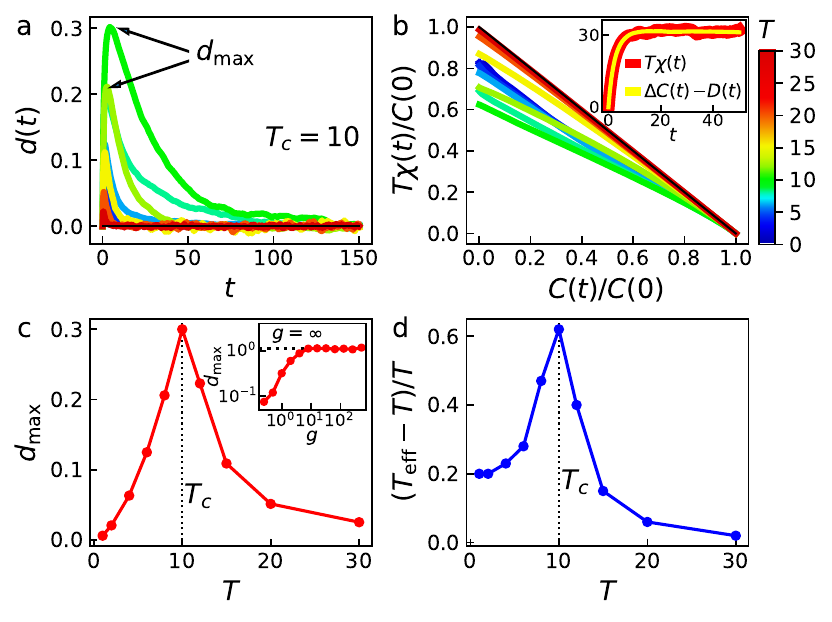}\\
\caption{FDT Violation.
a) Deviation $d(t)$ as a function of time, for different values of the temperature. b) Integrated response vs.  correlation, for different values of the temperature; both quantities are normalized by $C(t=0)$.
Inset: integrated response computed  by perturbing the system (red points) or by using Eqs.~\eqref{viola} and \eqref{rviolationg} (blue points), for $T=15$. c) Maximum of the deviation function, $d_{\rm max}$, as a function of temperature. Inset: behavior of $d_{\rm max}$ as a function of $g$. d) Asymptotic off-equilibrium offset $T_{\rm eff}(\infty)/T-1$ as a function of temperature; the black line corresponds to equilibrium. The parameters of the simulation are $J=1.5, v_0=4, g=4,\rho=0.25, N=512$, with temperatures described by the color bar and $T_c=10$ (light green). }
\label{fig:violation}
\end{figure}

\noindent
{\bf FDT violations across the phase diagram.}
We now proceed to compute numerically the response functions and the deviations from equilibrium, and monitor them across the phase diagram of the model. We use a discretized version of Eqs.~(\ref{eom}), in $d=3$, following a standard Euler integration scheme. 
We select a value of density for which violations are clearly visible in the whole phase space, and then tune the noise strength to span from the disordered to the ordered region.  We focus initially on checking the validity of Eq.(\ref{viola}). To do so, we compute the response in two ways, either perturbing the system or using Eqs.~\eqref{rviolationg} and (\ref{viola}) in absence of perturbation.
In fact, it is numerically more convenient to look at the integrated version of Eq.~\eqref{rviolationg}, i.e. $T \chi_{\alpha \alpha}(t) =   [ C_{\alpha \alpha}(0) - C_{\alpha \alpha}(t)] -  D_{\alpha \alpha}(t)$, where $\chi_{\alpha \alpha}(t)=\int_0^t d\tau R_{\alpha \alpha}(\tau)$ is the integrated response  \cite{cugliandolo2011effective} and $D_{\alpha \alpha}(t) =\int_0^t d\tau d_{\alpha \alpha}(\tau)$. 
For $T>T_c$ all coordinates $\alpha$ are statistically equivalent; for $T<T_c$ we consider the longitudinal component, i.e. we choose $\alpha$ as the direction of the polarization (see App.~\ref{app:numerical} for details).
%
The result is displayed in the inset of Fig.\ref{fig:violation}b, where we can clearly see that the response curves obtained in the two ways coincide within statistical fluctuations. Once convinced that we can trust Eq.(\ref{viola}), we can then exploit this expression to compute the deviation function, and therefore the response, without actually perturbing the system. 
The  deviation function $d(t)$ is displayed in Fig.\ref{fig:violation}a for several values of the temperature (we drop the coordinate index for notation convenience). In general, $d(t)$ shows a first quick increase on short timescales towards a maximum, and a slower decay to zero at longer times. We shall comment later on the origins of such dynamical behavior. For the time being, we notice that the violation is larger and more extended in time as the temperature approaches the critical value $T_c$  of the kinetic transition. 
 A different way - which is standard in the literature  - to capture out-of-equilibrium  behavior is to plot the integrated response parametrically as a function of the correlation function \cite{bouchaud1998spin}.
In equilibrium dynamics the expected behaviour is a straight line with slope given by (minus) the inverse temperature. On the contrary, as shown in Fig.\ref{fig:violation}b, in our model the curves clearly depart from the equilibrium one, and such departure looks more pronounced close to criticality, consistently with Fig.\ref{fig:violation}a.

To better investigate the role of temperature, we now introduce two synthetic quantifiers of the departure from equilibrium.
As a first measure, we consider the maximum $d_{\rm max}$ of the deviation function $d(t)$, which occurs on short timescales. In Fig.~\ref{fig:violation}c $d_{\rm max}$ is plotted as a function of temperature, and it displays a sharp maximum at the transition temperature. In the inset we also show $d_{\rm max}(T_c)$ as a function of $g$: as anticipated, it quickly converges to the limit $g\to \infty$ (i.e. standard Vicsek).

Another quantity that is widely used in the literature to quantify FDT violations is the so-called {\it effective temperature} \cite{hohenberg1989chaotic,cugliandolo1997energy,cugliandolo2011effective}, defined as the ratio between the correlation and the integrated response $T_{\rm eff}(t)=(C(0)-C(t))/\chi(t)$. Even though $T_{\rm eff}$ represents a real temperature only in a few cases, we will stick for convenience to this terminology. 
In Fig.~\ref{fig:violation}d we display the asymptotic value $(T_{\rm eff}(t\to \infty)-T)/T$ (which is zero at equilibrium) as a function of temperature. Also this quantity - which encodes FDT violations on long timescales - displays a maximum in correspondence of the critical temperature.

\begin{figure}[t] 
\includegraphics[width=0.45\textwidth]{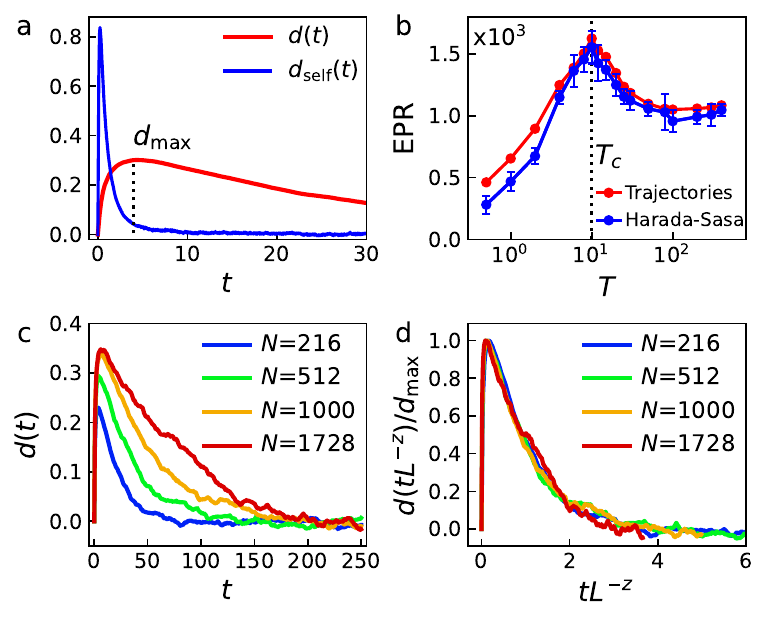} 
\caption{FDT violations across scales. a) Self-contribution $d_{\rm self}$ (red points) and full deviation (blue points) as a function of time, at $T=T_c$. b) EPR computed using the trajectories with Eq.~\eqref{epr} (red points) and the Harada-Sasa formula (blue points). c) Deviation curves $d(t)$ at $T=T_c(N)$ for several values of the system's size. d) Rescaled deviation $d(t)/d_{\rm max}(N)$ as a function of the rescaled time $t L^{-z}$, with 
$z=1.7$. The parameters of the simulation are $J=1.5, v_0=4, g=4,\rho=0.25, N=512$ (a and b). Results at $\rho=1.5$ can be found in App. \ref{app:numerical}
.}
\label{fig:scales}
\end{figure}

\vskip 0.15 cm
\noindent
{\bf FDT violations across scales}.
The above results indicate that FDT violations are maximal at the transition, and that the response is strongly out of equilibrium on all time-scales. The behavior in time of the deviation function $d(t)$, however, suggests that different processes might regulate the short time regime and the long time decay. In Fig.~\ref{fig:scales}a we plot the self contribution $d_{\rm self}=(1/N) \sum_i d_{ii}$ to the deviation and compare it to the full function. This contribution dominates the first regime of $d(t)$ and it  decays to zero on a timescale corresponding  to the location of $d_{\rm max}$.

Interestingly, $d_{\rm self}$ is connected to another quantifier of non-equilibrium,  the Entropy Production Rate (EPR), which measures  the amount of violation of detailed balance in the system.  The entropy production is defined as the logarithm of the ratio between the probability of a forward and a backward trajectory \cite{seifert2012stochastic,ritort2008nonequilibrium}, and it  can be computed  from the Onsager-Machlup action. We get 
\begin{equation}
{\rm EPR} = \lim_{\tau\to\infty} \frac{1}{\tau}\Big \langle \frac{J}{2T} \int_0^\tau d\tau \sum_{ij} \frac{dn_{ij}}{dt}\circ \sum_\alpha (v^\alpha_i-v^\alpha_j)^2\Big \rangle
\label{epr}
\end{equation}
where $\circ$ denotes the Stratonovitch prescription (see App.\ref{app:EPR}).
A discretized version of this expression can be used to evaluate the EPR from numerical trajectories. The result is displayed in Fig.\ref{fig:scales}b, and it shows that the EPR also peaks at the critical temperature \cite{ferretti2022signatures,yu2022energy}. In fact, the EPR and FDT violations are directly related to each other via Harada-Sasa relationships \cite{harada2005equality}, which we explicitly derive in App. \ref{app:HS},
\begin{align}
{\rm EPR} &= 
 \sum_{i,\alpha} \int \frac{d \omega}{2 \pi}  \frac{\omega}{T}
\left[ \omega\tilde{C}_{ii}^{\alpha\alpha}(\omega) - 2 T  \mathfrak{Im} (\tilde{R}_{ii}^{\alpha\alpha}(\omega)) 
\right] \nonumber \\
&=\frac{N}{T} \frac{ d}{dt} \Big ( \sum_\alpha d_{\rm self}^{\alpha \alpha}(t)\Big ) \Big |_{t=0^+}
\label{HS}
\end{align}
We recognize in the brackets the Fourier transform of the elements $d^{\alpha \alpha}_{ii}(t)$ of the deviation matrix, which also contribute to $d_{\rm self}$. We have verified this relation, by computing  Eq.~(\ref{HS}) via Mallivian weight sampling \cite{warren2012malliavin} and comparing the result with Eq.~(\ref{epr}), see Fig.\ref{fig:scales}b and App.~\ref{app:HS}.

 The connection with the EPR helps understanding the origin of the short time regime of the FDT violation: it is directly determined by the production of entropy in the system. Eq.(\ref{epr}) shows that the EPR depends on the rate of change of the interaction network $dn_{ij}/dt$, that is on how quickly nearby particles enter and exit their interaction range. It is therefore a local process. 
 Besides, since mutual diffusion in space is maximal at the transition 
 \cite{ferretti2022signatures,tu1998sound,chate2008modeling} 
 the EPR peaks at $T=T_c$ \footnote{At least for not too small values of $g$. We remind that $g$ regulates the amplitude of speed fluctuations. If it is too small speeds are not confined and they can alter the scenario so far described. This is however an unrealistic regime, which is not of interest to our analysis.}.

As we already noticed, strong deviations from equilibrium  are observed in the response also on the long timescales, as testified by the slow decay of the function $d(t)$ and by the behavior of $T_{\rm eff}(t=\infty)$ across the phase diagram (see Fig.~\ref{fig:violation}d).  
To explain this behavior, it is crucial to consider the interacting nature of the system and, in particular, the occurrence of strong cooperative phenomena as the transition point is approached. At large times, the deviation function is dominated by the non-self contribution, and it therefore involves correlations between pairs of particles. Close to $T_c$ such correlations become long-range, distributing off-equilibrium effects to the large scales and slowing down the decay of $d(t)$.

To confirm the collective nature of FDT violations at large times, we performed simulations for systems of different size $N$, at the critical temperature $T=T_c(N)$. The resulting $d(t)$ curves are displayed in Fig. \ref{fig:scales}c:  
the short time increase of the deviation function, being related to a local process, is always the same; 
on the contrary, the long time decay is slower the larger the size, a behavior typical of critical slowing down. Indeed, we show in Fig. \ref{fig:scales}d that the deviation curves obey dynamic scaling, falling one on top of the other once the time is rescaled by a characteristic time growing with the system's size as $L^z$. The value of the dynamic exponent used for the collapse is $z=1.7$, that is the value predicted by the incompressible field theory of the Vicsek model \cite{chen2015critical}. We conclude that - as long as the system is homogeneous enough in density and the incompressible regime holds \cite{dicarlo2022evidence} - long time FDT violations are a critical effect. Above a (density dependent) crossover size the cooperative processes at the transition are dominated by heterogeneities and phase separation \cite{chate2008collective,dicarlo2022evidence} and we expect the long time dynamics to be ruled by different laws. The simulations performed in this paper, as well as experiments on natural groups in $d=3$ \cite{cavagna2008new,cavagna2023natural}, all belong to the incompressible regime (see also App.\ref{app:numerical}.).

The general picture can be summarized as follows. Energy is injected in the system at the individual level, via the particle self-propulsion and the relative speed control function. Then, entropy is produced locally, every time two particles start/stop interacting with each other, and it is maximal at $T_c$ due to the fast network reshuffling. Violations of detailed balance immediately alter FDT relations on the short scales. The existence of long-range cooperative processes close to the transition then propagate off-equilibrium effects through the whole system. FDT deviations therefore persist at large times, and their overall amount peaks at the transition point.

\vskip 0.15 cm

\noindent
{\bf Conclusions.}
The way a system manages its energy budget is a crucial issue for living systems, which typically acquire external resources to build functional ordered structures and dissipate excess entropy in the environment. Understanding how this occurs and what determines the repartition of this process is therefore an important question \cite{gnesotto2018broken,lynn2022decomposing}. 
Indeed, it remains in general unclear whether activity in the individuals of an aggregation affects the dynamics also at the macroscopic level \cite{egolf2000equilibrium}; for example, recent analysis indicate that in motility induced phase separation equilibrium is restored at large scales \cite{nardini2017entropy,maggi2022critical}. Our results show that in fact in flocking systems non-equilibrium features do manifest on all scales, 
being stronger on the short times, and slowly decaying at long ones. The key ingredient for this to occur is the combination of  the individual driving enforcing self-propulsion, and the presence of an ordering transition coupling dynamical modes over the whole system.

Our analysis indicates that FDT violations give a more complete description of out-of equilibrium behavior than the EPR, as they capture violations on all scales, whereas the EPR is a strictly a local quantity. The different role of EPR and FDT violations has been pointed out in several works both at microscopic and at field theory level \cite{crisanti2012nonequilibrium, gnesotto2018broken,nardini2017entropy,caballero2020stealth}, but it becomes prominent in strongly interacting systems as the one considered here. In some cases, it might be useful to define a coarse-grained EPR \cite{esposito2012stochastic} capturing dissipation at different scales. 
In this way, an inverse dissipation scaling similar to what described here has been recently observed in self-similar reaction networks \cite{yu2021inverse}, self-similarity in the network structure playing the role of the scale-free behavior at the critical point. 

Finally, we note that our results explain what is observed in experiments on natural flocks and swarms: while flocks of birds behave as quasi-equilibrium systems \cite{mora2016local}, swarms of midges exhibit strongly out of equilibrium features \cite{cavagna2023natural}. This is consistent with our findings: flocks are highly polarized groups and therefore belong to a region where both EPR and FDT violations are small (see Fig.~\ref{fig:violation}); swarms, on the contrary, are quasi-critical systems and obey dynamic scaling \cite{attanasi2014finite,cavagna2017swarm} thereby representing instances at the peak of the violation-noise diagram.

\begin{acknowledgments}
This work was supported by MIUR grant PRIN-2020PFCXPE and by ERC grant RG.BIO (n. 785932).  We thank A. Cavagna and J. Crist{\'i}n for many interesting discussions. 
\end{acknowledgments}

\bibliography{COBBS-classic-bibliography}
\vfill\eject
\onecolumngrid 

\appendix
\section{Numerical simulations}\label{app:numerical}
\input{numericalinfo}

\section{Response function from the OM action} \label{app:OM}
\input{OM}

\section{Entropy Production Rate}\label{app:EPR}
\input{EPR}
\section{Harada-Sasa relationships}\label{app:HS}
\input{HaradaSasa}

\section{Equilibrium limits}\label{app:equilibrium}
\input{quasi_equilibrium}



\end{document}

%% file: numericalinfo.tex
The equations of motions for our model are, 
\begin{equation}
  \dfrac{d{\mathbf{r}}_i}{dt} =\mathbf{v}_i,\quad\quad\quad\quad
    \dfrac{d{\mathbf{v}}_i}{dt} =- J\sum_j n_{ij}(t) (\mathbf{v}_i - \mathbf{v}_j ) - g(|\mathbf{v}_i| - v_0) \dfrac{\mathbf{v}_i}{|\mathbf{v}_i |} + \boldsymbol{\xi}_i \ ,
\label{stochastic_diff_eq_explicit}
\end{equation}
where $\boldsymbol{\xi}_i $ is a Gaussian white noise with variance $\langle  \xi_i^\alpha (t) \ \xi_j^\beta (t') \rangle = 2 T \delta_{ij} \delta_{\alpha \beta} \delta(t-t')$. We simulated a discretized version of these equations, using an Euler discretization algorithm with $\Delta t = 1.0 \times 10^{-3}$, and computing at each time step the adjacency matrix $n_{ij}(t)=n_{ij}(| \mathbf{r}_i - \mathbf{r}_j|)$ according to a metric rule,
\begin{align}
\begin{cases}
   n_{ij}=1 \text{ }  \text{if } | \mathbf{r}_i - \mathbf{r}_j| \leq r_c  \\
    n_{ij}=0 \text{ }  \text{if } | \mathbf{r}_i - \mathbf{r}_j| > r_c  
    \end{cases}
\end{align}
All simulations have been performed with $r_c=1.0$. 

\subsection{Phase diagram}

\begin{figure}[t]
\centering
\includegraphics[width=0.5\textwidth]{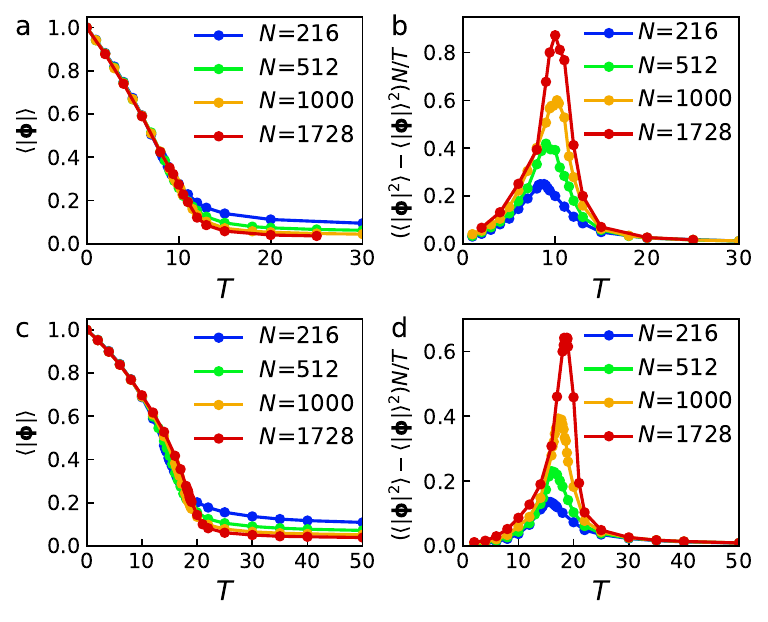}
\caption{Phase diagram of the model. a) Scalar polarization as a function of temperature, for different values of the system size. b) Fluctuations of the scalar polarization as a function of temperature, for different values of the system size. The fluctuations have been multiplied by a factor $N/T$ to mimic what in equilibrium systems represents the static susceptibility. The peak indicates a kinetic transition at approximately $T_c=10$. The parameters of the simulation are $v_0=4$, $J=1.5$, $g=4$ and $\rho=0.25$. Panels c) and d): same as a) and b) for the second set of parameters: $v_0=4$, $J=0.5$, $g=4$ and $\rho=1.5$. In this case the ordering transition occurs at a larger temperature.}
\label{fig:phase_diagram}
\end{figure}

After a preliminary exploration of the space of parameters, we focused on a few sets for which non-equilibrium effects are clearly visible in the whole phase diagram. Most of the simulations discussed in the main text are performed with $v_0=4$, $J=1.5$, $g=4$ and density $\rho=0.25$.
In Fig. \ref{fig:phase_diagram}a,b we report, for this set of parameters, the scalar polarization $|{\boldsymbol \Phi} |=|(1/N) \sum_i \bv{v}_i/|\bv{v}_i |$ (i.e. the order parameter) and its fluctuations as a function of temperature, for various system sizes $N$. The fluctuations have been multiplied by a factor $N/T$ to mimic what in equilibrium systems represents the static susceptibility. As it can be seen from the figure, the model exhibits the typical behavior of the Vicsek class, with a kinetic phase transition from a disordered state at large noise to a state of collective motion at low temperatures. The peak of the susceptibility at a given value of $N$ can be used to identify the critical temperature $T_c(N)$ at that size.

To check the role of density, we also performed simulations at a larger value of the density, with a second set of parameters: $v_0=4$, $J=0.5$, $g=4$ and $\rho=1.5$ (the smaller value of $J$ ensures that the transition temperature is not too large). The phase diagram for this set is displayed in Fig. \ref{fig:phase_diagram}c,d. The qualitative picture remains the same, but with a larger transition temperature.

\subsection{Density fluctuations and scaling}

The behavior shown in Fig. \ref{fig:phase_diagram} is consistent with a second-order transition scenario. However, Vicsek models with metric interactions are known to asymptotically exhibit a first-order transition, dominated by phase separation \cite{chate2008collective}. Therefore, for large enough sizes we expect density heterogeneities to become relevant and change the phenomenology at the critical point. More precisely, it can be shown \cite{dicarlo2022evidence}  that there is a crossover length-scale $\cal{R}$, whose value depends on the microscopic parameters of the model, such that for $L<{\cal R}$ density fluctuations are not strong enough and the large scale physics of the system is described by the incompressible field theory \cite{chen2015critical}, while for $L>{\cal R}$ density fluctuations become relevant and the incompressible regime does not hold anymore. Our numerical results indicate that - for the sizes that we simulated - the system is still in the homogeneous - incompressible regime.

\begin{figure}[b]
\centering
\includegraphics[width=0.9\textwidth]{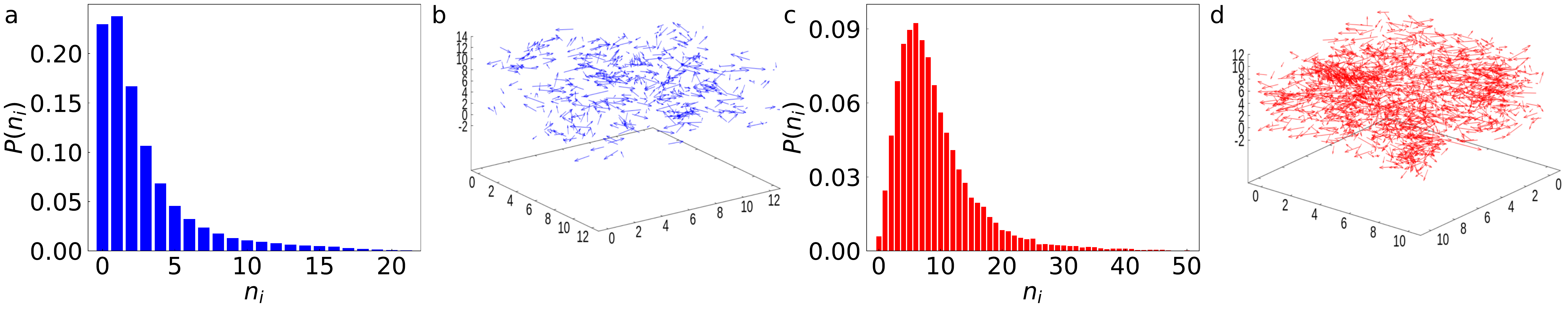}\  \quad
\caption{a) Histogram of the number of interacting neighbors $n_c$, for $\rho=0.25$ and $N=512$. b) Snapshot of a configuration in the stationary regime. c) and d) same as a) and b) but with $\rho=0.5$ and $N=1728$.}
\label{fig:hetero}
\end{figure}

To better investigate the role of heterogeneities we monitored the distribution of the number of interacting neighbors $n_i$ during the simulation. In Fig. \ref{fig:hetero}a,b we report this quantity close to the critical temperature for $\rho=0.25$ and $N=512$, the values used for most simulations, together with a snapshot of a configuration in the stationary regime. The histogram is well-peaked with small tails, indicating the absence of heterogeneities, as also visually confirmed from panel b.  For this value of the density, heterogeneities start to be seen at the largest size  considered $N=1728$; however, they are not yet strong enough to alter the large scale behavior. Indeed, as shown in the main text, dynamic scaling of the deviation curves is very nicely obeyed with the critical exponent $z=1.7$  predicted by the incompressible field theory. We also verified that the standard velocity correlation functions  obey dynamic scaling with the same exponent.

We note that $\rho=0.25$ corresponds to an average number of interacting neighbors of just a few individuals (see Fig. \ref{fig:hetero}a). We expect that increasing the density, such number should also increase making density fluctuations relatively less important. This is confirmed by simulations performed at larger densities. For $\rho=1.5$ (corresponding to an average $n_i$ of $9$ individuals) the system remains quite homogeneous also at the largest size considered, as reported in Fig. \ref{fig:hetero}c,d.  In Fig.~\ref{fig:scaling_rho1,5}  we show that - again - dynamic scaling is very well obeyed with $z=1.7$. Our results indicate that increasing the density makes the crossover length-scale ${\cal R}$ larger, meaning that a second-order like scenario and the incompressible field theory fixed point describe the large scale behavior up to larger system sizes.

\begin{figure}[t]
\centering
\includegraphics[width=0.6\textwidth]{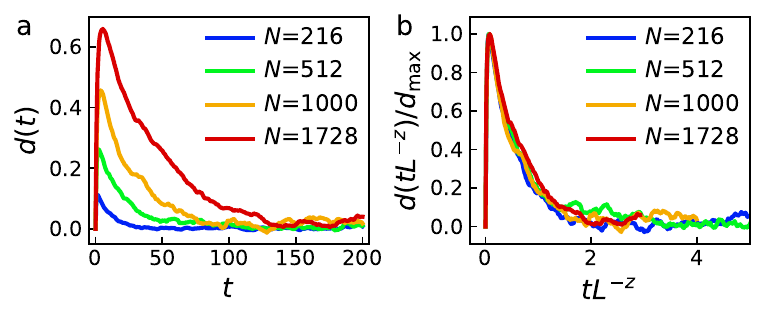}
\caption{a) Deviation curves $d(t)$ at $T=T_c(N)$ for several values of the system size. b) Rescaled deviation $d(t)/d_{\rm max}(N)$ as a function of the rescaled time $t L^{-z}$, with 
$z=1.7$. The parameters of the simulation are $v_0=4, J=0.5, g=4,\rho=1.5$.}
\label{fig:scaling_rho1,5}
\end{figure}

\subsection{Computing correlations and responses above and below $T_c$}
To compute correlations and response functions we followed the general definitions given in the main text. As already mentioned, it is numerically more convenient to look at the integrated response  $ \chi_{\alpha \beta}(t) =  \int_0^t d\tau R_{\alpha \beta}(\tau)$. 
In linear response theory, $\chi$ is directly related to the change of the mean group velocity under a constant perturbation, i.e. for ${\bv h}(t) = {\bv h} \ \theta(t)$ one has $V_\alpha(t)-V_\alpha(0)=\sum_\beta\chi_{\alpha \beta}(t) h_\beta$. Therefore, it can be easily computed from data by monitoring the mean velocity only. Since the probability distribution is symmetric under spacial reflections, the off-diagonal components of the response (and correlation) matrices are zero by symmetry. For $T>T_c$ the three diagonal components $R_{\alpha \alpha}$ (and $\chi_{\alpha \alpha}$) are equal. However, for $T<T_c$ the longitudinal component along the direction of the spontaneous polarization vector is different from the perpendicular one (exactly as in equilibrium alignment models), and they must be treated independently. In our analysis we focus on the longitudinal response.

Measuring global correlations and responses in the ordered phase has additional difficulties related to the finite size of the system and the so-called {\it wandering} of the order parameter. Let us now discuss this issue.
A key feature of alignment models is that they exhibit, at low temperatures,  a spontaneous symmetry breaking in the thermodynamic limit. When the system polarizes at the ordering transition, giving rise to a state of collective motion, one specific direction is selected out of the many equivalent ones consistent with the rotational symmetry of velocity alignment. A remnant of this continuous symmetry for $T<T_c$ is the presence of soft modes (the so-called Goldstone modes)  in the manifold orthogonal to the global polarization, which makes perpendicular fluctuations stronger than longitudinal ones \cite{brezin1973feynman,brezin1973critical}. For finite sizes, ergodicity is not strictly broken and the system can actually dynamically explore this manifold, slowly changing the direction of the global polarization: this is what is usually referred to as the `wandering' of the order parameter. This diffusive rotational motion occurs with diffusion coefficient of order $1/N$ and it becomes visible on numerical simulations on finite systems for large enough times,  making extremely difficult to compute reliably the {\it bona fide} correlation functions of the vectorial order parameter. 

\begin{figure}[t]
\centering
\includegraphics[width=0.3\textwidth]{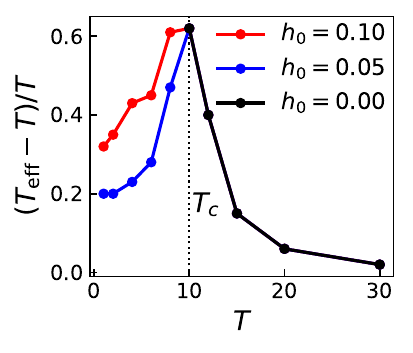}\ 
\caption{Effective temperature $T_{\rm eff}(\infty)$ as a function of temperature; the relative difference with respect to the equilibrium value is reported on the $y$-axis. The parameters of the simulation are $J=1.5, g=4,\rho=0.25, N=512$. 
The low temperature branch $T<T_c$ has been computed with two different values of the external field.
}
\label{fig:T_eff_appendice}
\end{figure}

To address this issue, for $T<T_c$ we explicitly break the symmetry with a small external field $h_0$ to make sure the order parameter fluctuates around this direction without wandering away. Then, we compute all the relevant quantities (correlations, responses, deviation functions) in presence of the field, and measure the off-equilibrium indicators. Finally, we verify that the results are robust upon decreasing the field. We note that this procedure (computing at finite $h$ and then sending the field to zero) is also used in theoretical equilibrium studies of Heisenberg models \cite{brezin1973feynman}. More precisely, we performed simulations for two different values of the external field $h_0=0.10$ and $h_0=0.05$. The external field explicitly breaks the symmetry along the direction $\hat{x}$, and correlation and response functions are computed for the $x$ component of the vectorial order parameter. Fig.~\ref{fig:T_eff_appendice} shows the deviation from equilibrium encapsulated by the effective temperature, as defined in the main text. The low-temperature phase branch depends on the value of $h_0$. However, reducing $h_0$ lowers the obtained effective temperature. 
Hence, the trend strongly suggests that, in the (computationally hard) limit $h_0 \to 0$ the peak at the transition point becomes even sharper, thus supporting the discussion and the general picture described in the main text.

Finally, we notice that the FDT violation seems to approach a non-zero limit for $T\to 0$, contrary to what happens for the EPR  (Fig~\ref{fig:violation})c). This behavior is related to the fact that, while the EPR is computed in zero field, the FDT violation below $T_c$ is evaluated (for the reasons discussed above) with the small field $h_0$. The external field, in presence of periodic boundary conditions, introduces an additional probability current and consequent FDT violations, not explicitly considered in the EPR formula (see \cite{fischer2018large} for a similar case). However, this effect decreases with $h_0$ (as also shown in Fig.~\ref{fig:T_eff_appendice}) and it therefore vanishes when $h_0\to 0$.

%% file: OM.tex

The Onsager-Machlup (OM) formalism is a well-known path-integral approach to Langevin stochastic dynamics that allows computing expectation values starting from the dynamical equations. In this section we briefly describe the method, as applied to our case. More details can be found in standard textbooks of stochastic processes.

Let us rewrite the equations of motion 
(Eq.~\eqref{eom} of the main text) in a more compact form
\begin{equation}
 \dfrac{d\bv{v}_i}{dt} =\mathbf{F}_i(\mathbf{v},\bv{r}) + \boldsymbol{\xi}_i \ \quad \quad \frac{d\bv{r}_i}{dt}=\bv{v}_i(t)
\label{stochastic_diff_eq}
\end{equation}
where $\mathbf{F}_i=-\partial {\cal H}/\partial \bv{v}_i$, $\bv{v}(t)\equiv\{\bv{v}_i(t)\}$, $\bv{r}(t)\equiv\{\bv{r}_i(t)\}$, and $\boldsymbol{\xi}(t) \equiv \{ \boldsymbol{\xi}_i(t)\}$ - we remind - is a Gaussian white noise with variance $\langle \xi_i^\alpha(t) \xi_j^\beta(t') \rangle = 2 T \delta_{\alpha \beta} \delta_{ij}\delta(t-t')$.
 The main idea behind the OM approach is that any physical observable $O(t)$ at time $t$ is a random variable, which can only depend on the noise realization for times before $t$, and on the initial conditions. Its average value $\langle O(t) \rangle$ is therefore obtained by averaging over the functional distribution of the noise realizations. However, since observables are ultimately functions of velocities and positions, such averages can be conveniently re-expressed in terms of the distribution probability of the velocities and positions themselves. Given a thermal history $\boldsymbol{\xi}(t)$, velocities are obtained via Eq.~\ref{stochastic_diff_eq} as $\bv{v}_{i\boldsymbol{\xi}}(t)= \bv{v}_{i}(0) + \int_0^t dt^\prime [ \boldsymbol{F}_i(\bv{v},\bv{r}) + \boldsymbol{\xi}_i ]$. Positions are instead deterministically obtained from the velocities, i.e. $\bv{r}_{i\bv{v}}(t)= \bv{r}_{i}(0) + \int_0^t dt^\prime \bv{v}_i(t^\prime)$.
 The distribution of velocities and positions, given the initial conditions, can therefore be immediately derived as
 \begin{equation}
    P\left [ \bv{v }(t) , \bv{r }(t) |  \bv{v }(0) ,  \bv{r}(0)\right ] = \int  D \left[ \boldsymbol{\xi }(t')\right] P\left [ \boldsymbol{\xi }(t') \right ]\prod_i\delta\left[ \bv{v}_i(t') - \bv{v}_{i\boldsymbol{\xi }}(t') \right]  \ \delta [ \bv{r}_i(t) - \bv{r}_{i\bv{v}}(t) ].  
\end{equation}
 where thermal histories are considered between times $0$ and $t$.
Inverting the dependence between $\bv{v}(t)$ and $\boldsymbol{\xi }(t)$, and exploiting the explicit form of the Gaussian noise distribution, we get
\begin{equation}
\begin{aligned}
     P\left [ \bv{v }(t) , \bv{r }(t) |  \bv{v }(0) ,  \bv{r}(0)\right ] 
    = & e^{ - \large \int_0^t dt' \left [ \frac{1}{4T}   \sum_i \left( \Dot{\bv{v }}_i - \boldsymbol{F}_i(\bv{v},\bv{r}) \right)^2 +f[\bv{v},\bv{r}] \right ] }  \delta [ \bv{r}_i(t) - \bv{r}_{i\bv{v}}(t) ]\\
    =  &e^{-S[\bv{v}(t),\bv{r}(t)|\bv{v}(0),\bv{r}(0) ]} \delta [ \bv{r}_i(t) - \bv{r}_{i\bv{v}}(t) ]\ .
\end{aligned}
\label{OMaction}
\end{equation}
The term $S[\bv{v}(t)),\bv{r}(t)|\bv{v}(0),\bv{r}(0)]$ is called ``Onsager-Machlup" (OM) action. The function $f[\bv{v},\bv{r}]$ comes from the Jacobian factor $|{\rm det} (\delta \boldsymbol{\xi}/\delta \bv{v})|$ and its explicit expression depends on the stochastic integration scheme. Using Ito's rules of calculus (the more convenient ones for simulations) the Jacobian is equal to $1$ and $f=0$. 
The expected value of any observable over the thermal history can be computed as, 
\begin{equation}
    \langle O(t) \rangle = \int D \left[ \boldsymbol{v }(t')\right]  D \left[ \boldsymbol{r }(t')\right]   O(t^\prime)  e^{-S[\bv{v}(t'),\bv{r}(t^\prime)|\bv{v}(0),\bv{r}(0)]} \delta [ \bv{r}_i(t') - \bv{r}_{i\bv{v}}(t') ] \ ,
    \label{OMexp}
\end{equation}
with
\begin{equation}
S[\bv{v}(t),\bv{r}(t)|\bv{v}(0),\bv{r}(0)] = \frac{1}{4T} \sum_{i,\alpha} \int_0^t dt' \  \left(\dot{v}_i^\alpha (t') - F_i^\alpha(t')  \right)^2
\label{OMS}
\end{equation}
Expression \eqref{OMexp} can be eventually averaged over the distribution of the initial conditions.  
In the main text, for notational convenience, we omitted the dependence on the initial conditions, and included the delta function on the positions in the measure.
We note however that in the stationary state, if the system is ergodic, the dependence on initial conditions is lost.

\subsection{The response function}
Using Eq.~\eqref{OMexp} we can immediately compute the response function, i.e.
\begin{equation}
\begin{aligned}
R^{\alpha \beta}_{ij}(t,s)= & \dfrac{ \delta \langle v_i^\alpha(t) \rangle_h } {\delta h_j^\beta(s) } \bigg|_{h=0} = - \Bigg \langle v_i^\alpha(t) \dfrac{\delta S}{\delta h_j^\beta(s)} \Bigg  \rangle \Bigg|_{h=0} \\
=& \frac{1}{2T} \langle v_i^\alpha(t)(\dot{v}_j^\beta (s) - F_j^\beta(s))\rangle  =
 \frac{1}{2T}  \dfrac{\partial C_{ij}^{\alpha\beta}(t,s)}{\partial s} - \frac{1}{2T}\langle v_i^\alpha(t)F_j^\beta(s))\rangle \ ,
\label{risposta_non_stazionaria}
\end{aligned}
\end{equation}
where, in the last line, averages are performed in zero field. In the stationary state, the response and correlation function depend on $t$ and $s$ only through their difference. A simpler expression for the response function can then be obtained. Let's assume without loss of generality that $t>s$. In this case, by causality, $R^{\alpha\beta}_{i j}(s-t)=0$. By subtracting it from Eq.~\eqref{risposta_non_stazionaria} we then get, for $t>s$,
\begin{equation}
      T  R^{\alpha\beta}_{i j}(t-s) = - \dfrac{1}{2} \left[ \dfrac{d C_{ij}^{\alpha\beta}(t-s)}{dt} + \dfrac{d C_{ij}^{\alpha\beta}(s-t)}{dt} \right ]-d_{ij}^{\alpha\beta}(t-s) \ ,
\label{risposta_sr}
\end{equation}
where $d_{ij}^{\alpha\beta}(t-s)=(1/2) \langle v_i^\alpha(t)F_j^\beta(s)-v_i^\alpha(s)F_j^\beta(t) \rangle$. The off-diagonal terms with $\alpha\ne\beta$ are  identically zero by symmetry arguments. For the diagonal ones we get,
\begin{equation}
      T  R^{\alpha\alpha}_{i j}(t-s) = - \dfrac{d C_{ij}^{\alpha\alpha}(t-s)}{dt} -d_{ij}^{\alpha\alpha}(t-s) \ ,
      \label{response-mt}
\end{equation}      
which is Eq.~\eqref{rviolation} of the main text. An explicit computation of the term $d_{ij}^{\alpha\alpha}(t-s)$ for the model at hand can be obtained using the expression of the force
\begin{equation} 
{\bv F}_i = -J\sum_{j} n_{ij} ({\bv v}_i - {\bv v}_j) - g (|{\bv v}_i| -v_0) \frac{{\bv v}_i}{|{\bv v}_i|} \ ,
\label{force}
\end{equation}
and it gives,
\begin{equation}
    \begin{aligned}
    d_{ij}^{\alpha\alpha}(t-s)  = &\dfrac{J}{2}\sum_k  \langle v_i^\alpha(s) n_{jk}(t) (v_j^\alpha(t)-v_k^\alpha(t)) \rangle - \dfrac{J}{2}\sum_k  \langle v_i^\alpha(t)n_{jk}(s)(v_j^\alpha(s)-v_k^\alpha(s)) \rangle \\
\quad\quad & +    \dfrac{gv_0}{2} \left(\langle  v_i^\alpha(t)\dfrac{ v_j^\alpha(s)}{|\boldsymbol v_j(s)|} \rangle -\langle  v_i^\alpha(s)\dfrac{ v_j^\alpha(t)}{|\boldsymbol v_j(t)|} \rangle \right)
\end{aligned}
\label{d_ij_alpha}
\end{equation}
To obtain the global response to a homogeneous field (see main text), we sum both sides of Eq.~\eqref{response-mt} over the $i$ and $j$ indexes and divide by $N$. We then get,
\begin{equation}
T R_{\alpha\alpha}(t-s) = - \frac{d}{dt} C_{\alpha \alpha}(t-s) - d_{\alpha\alpha}(t-s)
\end{equation}
where $C_{\alpha \alpha} (t-s)= N \langle V_\alpha(t) V_\alpha(s) \rangle$ is the correlation of the global velocity and $d_{\alpha\alpha}(t-s)$ can be computed directly from Eq.~\eqref{d_ij_alpha}. Interestingly, given that the matrix $n_{ij}$ is symmetric, many terms simplify and we get
\begin{equation}
    d_{\alpha\alpha}(t-s) = \frac{1}{N} \sum_{ij} d_{ij}^{\alpha\alpha} (t-s)= 
    \dfrac{gv_0 N}{2}  \left[ \langle V_{\alpha}(t) \Phi_{\alpha}(s) \rangle - \langle V_{\alpha}(s) \Phi_{\alpha}(t) \rangle\right],
\end{equation}
recovering Eq.~\eqref{viola} of the main text. 
%
%

%% file: EPR.tex
The average entropy production $\Sigma(t)$ can be defined as the averaged logarithm of the ratio of the probability of observing a trajectory of length $t$, and the probability of the reversed process \cite{lebowitz1999gallavotti,seifert2012stochastic}. For a Markov process,  it can be conveniently expressed in terms of the OM action, 
\begin{equation}
\begin{aligned}
\Sigma(t) &= \Bigg \langle \log \dfrac{P[{\bv v}(t),\bv{r}(t)]}{P[{\bv v}^\dagger(t),\bv{r}^\dagger(t)]} \Bigg \rangle =
 \Bigg \langle \log \dfrac{P[{\bv v}(t),\bv{r}(t) | {\bv v}(0),\bv{r}(0)] P_0({\bv v}(0),\bv{r}(0))}{P[{\bv v}^\dagger(t),\bv{r}^\dagger(t) | {\bv v}^\dagger(0), \bv{r}^\dagger(0)] P_0({\bv v}^\dagger(0),\bv{r}^\dagger(0)) }\Bigg  \rangle \\
 & = \langle S[{\bv v}^\dagger(t),\bv{r}^\dagger(t) | {\bv v}^\dagger(0),\bv{r}^\dagger(0)] -  S[{\bv v}(t),\bv{r}(t) | {\bv v}(0),\bv{r}(0)] \rangle + \Bigg \langle \log \dfrac{P_0({\bv v}(0),\bv{r}(0))}{P_0({\bv v}^\dagger(0),\bv{r}^\dagger(0))  } \Bigg \rangle
\end{aligned}
\label{EP}
\end{equation}
where averages are taken over the ensemble of the stochastic paths. The last term in the r.h.s of Eq.~\eqref{EP}  is a boundary term that does not depend on time:  since we are ultimately going to take the temporal average rate, we disregard it in the following. The first two terms at the r.h.s. can be computed from the explicit expression of the OM action. The state variables transform under time reversal as $\boldsymbol{r}(t')\mapsto \boldsymbol{r}^\dagger(t') = \boldsymbol{r}(t -t')$ and ${\bv v}(t')\mapsto{\bv v}^\dagger(t') = - {\bv v}(t-t')$.  For what concerns the forces ${\bv F_i(t')}$  (see Eq.\eqref{force}), given that positions enter only via the adjacency matrix $n_{ij}$ (which is invariant under time reversal),  we have ${\bv F_i(t')} \mapsto{\bv F}_i^\dagger(t')= - {\bv F_i(t-t')}$. From Eq.\eqref{OMS} we then get 
\begin{equation}
\begin{aligned}
S[{\bv v}^\dagger(t),\bv{r}^\dagger(t)| {\bv v}^\dagger(0),\bv{r}^\dagger(0)] -  S[{\bv v}(t),\bv{r}(t) | {\bv v}(0),\bv{r}(0)] & =  \frac{1}{4T}\sum_i\sum_\alpha \int_0^{t} dt \left[ \left (\frac{dv_i^{\alpha}}{dt} + F_i^\alpha \right )^2 -\left (\frac{dv_i^{\alpha}}{dt} - F_i^\alpha\right )^2 \right]= \\
& = \frac{1}{T}\sum_i\sum_\alpha \int_0^{t} dt  \  \frac{dv_i^{\alpha}}{dt} \circ F_i^\alpha  \ ,
\end{aligned}
\end{equation}
where $\circ$ denotes the Stratonovich prescription, always arising regardless of the type of stochastic calculus in the original integral (indeed the inverse process always has by construction the complementary prescription of the forward one).

We can now evaluate the average entropy production rate.  In the stationary state, and  assuming ergodicity, we get
\begin{equation}
\begin{aligned}
{\rm EPR}   &=  \lim_{t\to\infty}\frac{ \Sigma(t) }{t}  =   \lim_{t\to\infty}  \frac{1}{t}  \frac{1}{T} \sum_i\sum_\alpha  \int_0^{t} dt  \ \Big \langle \frac{dv_i^{\alpha}}{dt} \circ F_i^\alpha \Big \rangle \\
&= \frac{1}{T} \sum_{i}\sum_{\alpha}\Big \langle \frac{d v_i^{\alpha}}{dt} \circ F_i^{\alpha}\Big \rangle \\
&= -\frac{1}{T} \sum_{ij}\sum_{\alpha}\Big \langle \frac{d v_i^{\alpha}}{dt} \circ \left[J n_{ij}(v_i^\alpha-v_j^\alpha) + g \frac{v_i^{\alpha}}{|\boldsymbol{v}_i|}(|\boldsymbol{v}_i|-v_0)\right]\Big \rangle.
\label{heat-epr}
\end{aligned}
\end{equation}
 This expression can be further manipulated observing that, in the stationary state,
\begin{equation}
 0 = \Big \langle  \dfrac{d \mathcal{H}}{dt} \Big \rangle =\frac{J}{2}  \sum_{i,j}\sum_{\alpha} \Big \langle \frac{d n_{ij}}{dt}\circ(v_i^\alpha-v_j^\alpha)^2\Big \rangle  -\sum_i\sum_{\alpha} \Big \langle F_i^{\alpha} \circ \dfrac{dv_i^{\alpha}}{dt} \Big \rangle.
\end{equation}\label{primo principio}
Hence, substituting  in Eq.~\eqref{heat-epr}, we find
\begin{dmath}
{\rm EPR} = \frac{J}{2T}  \sum_{i,j} \Big \langle \frac{d n_{ij}}{dt}\circ\sum_{\alpha}(v_i^\alpha-v_j^\alpha)^2\Big \rangle.
\label{EPR-full-parti}
\end{dmath}

%% file: HaradaSasa.tex
\subsection{Harada-Sasa relationship from the OM action}

The Harada-Sasa relationship is a formula that connects the EPR with a violation of the FDT, firstly derived for Langevin dynamics in \cite{harada2005equality}.
To illustrate such relation in our case, let us consider the expression for the response function obtained in the previous section, and take the difference $R^{\alpha \alpha}_{ii}(t-s)-R^{\alpha \alpha}_{ii}(t-s)$. By causality only one of the two responses is different from zero and we get (without assuming any order between the two times)
\begin{equation}
    R^{\alpha\alpha}_{i i}(t-s) - R^{\alpha\alpha}_{i i}(s-t) = - \dfrac{1}{T}  \dfrac{d C_{ii}^{\alpha\alpha}(t-s)}{dt} -\dfrac{1}{T} d_{ii}^{\alpha\alpha}(t-s).
    \label{diff}
\end{equation}
We take now $s=t-\Delta t$, divide both sides by $\Delta t$, and take the limit as $\Delta t $ goes to $0$. We consider first the left-hand side, 
\begin{equation}
\begin{aligned}
    &\lim_{\Delta t \to 0}\dfrac{R^{\alpha\alpha}_{i i}(\Delta t) - R^{\alpha\alpha}_{i i}(-\Delta t)}{\Delta t} = \lim_{\Delta t \to 0} \int_{-\infty}^{\infty}\dfrac{d\omega}{2\pi} \Tilde{R}^{\alpha\alpha}_{i i}(\omega) \dfrac{e^{-i\omega \Delta t} - e^{i\omega \Delta t}}{\Delta t} =\\
    &= - \int_{-\infty}^{\infty}\dfrac{d\omega}{2\pi} 2i \omega \Tilde{R}^{\alpha\alpha}_{i i}(\omega) = \int_{-\infty}^{\infty}\dfrac{d\omega}{2\pi} 2\omega \ \mathfrak{Im} \left(\Tilde{R}^{\alpha\alpha}_{i i}(\omega)\right),
\end{aligned}
\end{equation}
where $\Tilde{R}^{\alpha\alpha}_{i i}(\omega)= \int_{-\infty}^{\infty} dt \exp{(i\omega t)}R^{\alpha \alpha}_{ii}(t)$ is the Fourier transform of the response, and $\mathfrak{Im}$ denotes the imaginary part (the only one surviving inside the integral, being the response
 a real function).
 The two terms at the r.h.s of Eq.~\eqref{diff} can be handled in a similar way,
 \begin{equation}
         \lim_{\Delta t \to 0} \frac{1}{\Delta t} \dfrac{1}{T}  \dfrac{d C_{ii}^{\alpha\alpha}(\Delta t)}{dt} = 
         -\dfrac{1}{T}\int_{-\infty}^{\infty}\dfrac{d\omega}{2\pi} \omega^2 \Tilde{C}_{ii}^{\alpha\alpha}(\omega), 
\end{equation}
and
\begin{equation}
    \begin{aligned}
     \lim_{\Delta t \to 0} \frac{1}{\Delta t}   \dfrac{1}{T} d^{\alpha\alpha}_{i i}(\Delta t) =& \dfrac{1}{T}\lim_{\Delta t \to 0}\dfrac{\langle v_i^{\alpha}(t) F_i^{\alpha}(t-\Delta t) \rangle - \langle v_i^{\alpha}(t- \Delta t) F_i^{\alpha}(t) \rangle}{2\Delta t} =\\
    =& \dfrac{1}{T}\lim_{\Delta t \to 0}\dfrac{\langle v_i^{\alpha}(t) F_i^{\alpha}(t-\frac{\Delta t}{2}) \rangle - \langle v_i^{\alpha}(t- \frac{\Delta t}{2}) F_i^{\alpha}(t) \rangle}{\Delta t} =\\
    =& \dfrac{1}{T}\lim_{\Delta t \to 0}\dfrac{\langle v_i^{\alpha}(t+\frac{\Delta t}{2} ) F_i^{\alpha}(t) \rangle - \langle v_i^{\alpha}(t- \frac{\Delta t}{2}) F_i^{\alpha}(t) \rangle}{\Delta t} = \\
    =&\dfrac{1}{T} \langle \Dot{v}_i^{\alpha} \circ  F_i^{\alpha}\rangle,
    \end{aligned}
\end{equation}
where stationarity has been used, together with the explicit definition of the Stratonovich differentiation.
Putting all the pieces together, we find,
\begin{equation}
    \int_{-\infty}^{\infty}\dfrac{d\omega}{2\pi} 2\omega \ \mathfrak{Im} \left(\Tilde{R}^{\alpha\alpha}_{i i}(\omega)\right) = \dfrac{1}{T}\int_{-\infty}^{\infty}\dfrac{d\omega}{2\pi} \omega^2 \Tilde{C}_{ii}^{\alpha\alpha}(\omega) - \dfrac{1}{T} \langle \Dot{v}_i^{\alpha} \circ  F_i^{\alpha}\rangle .
\end{equation}
Finally, comparing with Eq.~\eqref{heat-epr}, we get
\begin{equation}
    {\rm EPR} =   \sum_{i,\alpha} \int_{-\infty}^{\infty} \frac{d \omega}{2 \pi}  \frac{\omega}{T}
\left[ \omega\tilde{C}_{ii}^{\alpha\alpha}(\omega) - 2 T  \mathfrak{Im} (\tilde{R}_{ii}^{\alpha\alpha}(\omega))  
\right] \ ,
\label{EPR2}
\end{equation}
which is the Harada-Sasa formula, Eq.~\eqref{HS} of the main text. 

\subsection{Numerical check of the Harada-Sasa formula}
In the previous sections we derived two distinct expressions of the EPR. The first one, detailed in Sec.\ref{app:EPR}, can be used to compute the EPR directly from the trajectories from the numerical simulation, by implementing a discrete version of the time integral appearing in Eq.~\eqref{heat-epr}. This method is very efficient, it has been already tested in \cite{ferretti2022signatures}, and it produces the red curve in Fig.~\ref{fig:scales}b of the main text. The Harada-Sasa relation, on the other hand, is more complicated to check numerically, especially for strongly interacting systems as the one considered in this work. In this section we discuss why, and how we addressed the various technical issues.

To compute numerically the r.h.s. of Eq.~\eqref{EPR2}, one needs first of all to compute the local responses $R^{\alpha\alpha}_{ii}(t)$. As discussed in the main text, local responses are more demanding to calculate than global ones. Luckily, one can resort to the so-called Malliavin weight sampling  \cite{warren2013malliavin}. The basic idea of this method can be grasped by looking back at the derivation of Eq.~\eqref{risposta_non_stazionaria}. From the second line, we see that the response can be expressed as $R^{\alpha \alpha}_{ii}(t-s)=1/(2T) \langle v_i^\alpha(t) \xi_i^{\alpha}(s)\rangle$, where averages are performed in zero field. Therefore, by keeping track of the noise stories, one can compute local responses without perturbing the system. We refer the reader to the relevant literature for smart implementations of the method.

One issue, specifically related to polar systems of finite size, is the spurious effect due to the wandering of the order parameter below the ordering transition.
To contain the effect of wandering, we notice that FDT relations and the Harada-Sasa formula only involves derivatives of the correlation function, and they therefore remain the same either considering simple or connected correlations.  At low temperatures we then used connected correlation functions, \begin{equation}
    C_{ii}^{\alpha\alpha}(t-s) = \langle \delta v_i^{\alpha}(t)\delta v_i^{\alpha}(s) \rangle, \quad\quad  \delta v_i^{\alpha}(t)= v_i^{\alpha}(t) - \dfrac{1}{N}\sum_j v_j^{\alpha}(t) \ ,
\end{equation}
where the space average is used as a proxy of the average mean velocity. In the thermodynamic limit, this procedure exactly recovers the theoretical prediction Eq.~\eqref{EPR2}. At the same time, by automatically subtracting the wandering, it also provides a better estimate of this expression also for finite systems.
For consistency, and to ensure a better comparison, we also evaluate the EPR from  Eq.~\eqref{heat-epr} using velocities in the center of mass reference frame.


The Harada-Sasa relation involves the Fourier transforms of the correlation and response functions, and the multiplication of their difference by a (diverging) factor $\omega$. This implies that, whenever there is any small error in determining precisely these function, the integral of Eq. \eqref{EPR2} can differ significantly from the theoretical value, possibly even becoming divergent. The first trivial source of error is the finite time window of observation of the correlation and response functions, affecting the discrete resolution of the $\omega$ points, and thus the accuracy of the integral. Less trivial, instead, is the impact of the finite integration step $\Delta t$ of the numerical simulation: it is not commonly known that, even at equilibrium, the temperature defined as the noise variance coefficient and temperature defined as the factor appearing in the Gibbs-Boltzmann distribution are not the same for discrete-time dynamics, but they differ by a term of $O(\Delta t)$ \cite{cavagna2014dynamical}. As a consequence, even at equilibrium, discrete dynamics response differs from the continuum expression by a term  $O(\Delta t)$, which can affect even quite significantly the result of the integral. To address this problem, we hence performed simulations with a smaller $\Delta t = 1.0 \times 10^{-4}$, at the price of longer times needed for thermalization. 
Finally, the finite time resolution of the discrete points of the correlation functions imposes, in Fourier's space, a constraint on the tails of the transformed functions, making them hit zero at a certain cutoff; to avoid this problem, a high resolution in time is needed for the correlation and response functions, at the price of stronger noise when taking the derivative of the correlation function and the integrated response, because of smaller steps. All things considered, the signal is very noisy and it requires a lot of samples. The way we estimated the integral at the r.h.s. of Eq.~\eqref{EPR2}, and produced the blue curve of Fig. \ref{fig:scales}b, is the following: we performed many runs of numerical simulations, divided them into groups and computed the average correlation and response functions for each group. Then, for each group average, we computed a cut-offed version of the integral,
\begin{equation}
    F(\Omega) =   \sum_{i,\alpha} \int_{-\Omega}^{\Omega} \frac{d \omega}{2 \pi}  \frac{\omega}{T}
\left[ \omega\tilde{C}_{ii}^{\alpha\alpha}(\omega) - 2 T  \mathfrak{Im} (\tilde{R}_{ii}^{\alpha\alpha}(\omega)) \ .
\right]
\end{equation}
This function should have, was not for the noise, an asymptote for $\Omega \to \infty$. We then assign a value to the integral by performing a numerical fit of the asymptote. We repeat the procedure for each group  and we then take the average of them all. The error bars are computed as the standard deviation across groups and, considered the many sources of error, they are likely to be still underestimated. We also tried a different kind of analysis, where to reduce the impact of noise we first fit the integrand with appropriate functional forms, and then compute the integral.
The results are in general not very different for the average, while the associated  error bars are smaller, while realistically they should be bigger.
Finally, by comparing with the EPR computed from trajectories (see Fig.~\ref{fig:scales}b), we see that there is still a slight, but systematic, bias  below $T_c$ where the EPR via the Harada-Sasa formula is underestimated. We were not able to pinpoint the reason of this. Nevertheless, in light of all the aforementioned numerical challenges, we consider our results a quite satisfactory validation of the formula.

%% file: quasi_equilibrium.tex
\subsection{Recovering equilibrium in the active frame}

In the main text we derived an expression, Eq.~\eqref{viola}, where FDT violations are quantified in terms of simple correlation functions between the mean velocity and the polarization. One way to read Eq.~\eqref{viola} is that, out of equilibrium, the information content of the polarization is larger than the one of the collective velocity, the first influencing the future behavior of the second more than viceversa. In general, we expect violation of detailed balance to manifest differently in different degrees of freedom: by introducing fluctuating speeds we exploited this property to find easily accessible signatures of FDT violations. 

Interestingly, in our system  the slow critical modes (i.e. the polarization) are the ones most affected by detailed balance violations. On the contrary, the speeds - which are fast modes and mark the difference between ${\mathbf \Phi}$ and ${\bf V}$, are not. To illustrate this, we now consider the behavior of the mean velocity in a modified frame, i.e. when subtracting the mean ``active trajectory" $v_0{\mathbf \Phi}$ from the individual velocities. We therefore define the reduced mean velocity ${\tilde {\bf V}}= {\bf V} -v_0{\mathbf \Phi}$ and show that it responds as an equilibrium quantity when we perturb the system with an external field conjugated to the velocity. To do so, we first observe that, from the equation of motion of the single particles, an equation of motion for the global velocity can be derived,
\begin{equation}
\frac{d V_{\alpha}(t)}{dt}= -gV_{\alpha}(t) + gv_0 \Phi_{\alpha}(t) + \zeta_{\alpha}(t),    
\end{equation}
with $\langle \zeta_{\alpha}(t) \zeta_{\beta}(t')  \rangle = (2T/N)\delta_{\alpha\beta}\delta(t-t')$.
Starting from this equation we can compute the antisymmetric average $\langle \Phi_\alpha(s) dV_{\alpha}(t)/dt\rangle - \langle \Phi_\alpha(t) dV_{\alpha}(s)/ds\rangle$. In the stationary state, we get (for $t>s$)
\begin{equation}
2 \Big \langle \Phi_\alpha(s) \frac{dV_\alpha(t)}{dt}\Big \rangle=
- g \big [ \langle \Phi_\alpha(s) V_{\alpha}(t) \rangle - \langle \Phi_\alpha(t) V_{\alpha}(s)\rangle \big ]+ \langle \Phi_\alpha(t)\zeta_\alpha(s)\rangle \ .
\end{equation}
The first term in the r.h.s. is proportional to the deviation function and it can be re-expressed in terms of the correlation and the response of the mean velocity (see Eqs.~\eqref{viola} and \eqref{response-mt}). The second term is nothing else than  the response of the polarization, i.e. $2T \delta \langle \Phi_\alpha(t)\rangle /\delta h_\alpha(s)|_{h=0}$, as it can be deduced from Eq.~\eqref{risposta_non_stazionaria}. Reordering the terms, we therefore get
\begin{equation}
   \frac{\delta \langle \tilde{V}_\alpha(t)\rangle}{\delta h_\alpha(s)}\Bigg |_{h=0} = \dfrac{N}{T}\langle  \Tilde{V}_{\alpha}(t) \dfrac{d}{ds}V_{\alpha}(s)\rangle \ ,
\end{equation}
which is precisely the relation holding for equilibrium systems. We note that this result reminds the restoring of equilibrium relations in the Lagrangian frame described in systems of non-equilibrium colloidal and granular particles \cite{speck2006restoring,chetrite2008fluctuation,crisanti2012nonequilibrium}

\subsection{Pointers limit}
One can ask whether an equilibrium limit exists for the model of Eqs.\eqref{eom}\eqref{ham}. Since motility is the primary source of violation of detailed balance, one could think that merely sending $v_0$ to zero should recover equilibrium, and that $d(t)$ trivially vanishes because of the $v_0$ prefactor (see Eq.~\eqref{viola}). However, a finite speed not only determines the rearrangement of the interaction network (and the consequent entropy production), but also the length of the individual directional degrees of freedom.  If we just send $v_0\to 0$, particles slow down, but all velocities become smaller while their confining potential tends to a Gaussian theory, which would become problematic in terms of ordering. The limiting model would have a very different phenomenology than the original one. The proper way to define an equilibrium limit is more subtle, and requires focusing on rescaled variables. Let us define
\begin{equation}
    \boldsymbol{\sigma}_i=\dfrac{\bv{v}_i}{v_0}.
\end{equation}
In these new variables, the equations of motion become,
\begin{equation}
    \frac{d\boldsymbol{\sigma}_i}{dt} =  J\sum_j n_{ij}(\boldsymbol{\sigma}_i-\boldsymbol{\sigma}_j) - g (|\boldsymbol{\sigma}_i|-1)\dfrac{\boldsymbol{\sigma}_i}{|\boldsymbol{\sigma}_i|} + \boldsymbol{\Tilde{\xi}}_i \  , \quad \quad
     \frac{d \mathbf{r}_i}{dt} = v_0\boldsymbol{\sigma}_i
\label{eom_pointer}
\end{equation}
with $\langle \boldsymbol{ \Tilde{\xi}}_i(t) \cdot \boldsymbol{ \Tilde{\xi}}_j(t') \rangle = 2d (T/v_0^2) \delta_{ij} \delta(t-t')$.
If we now rescale the temperature, i.e. $T\to {\tilde T}=T/v_0^2$ (which amounts to considering a random force in the original equation that, as the deterministic ones, scales with the speed), the only remaining dependence on $v_0$ is in the equation of motion for the positions and, implicitly, in the interaction network $n_{ij}$.
When taking the limit $v_0 \to 0$, the only thing happening is that positions `freeze' and the network becomes fixed. The equation for the $\{\boldsymbol{\sigma}_i\}$ then becomes the relaxational dynamics of a Heisenberg model with a soft constraint on the spins,
thus recovering an equilibrium model. Looking at the off-equilibrium deviation $d(t)$ written in terms of the rescaled variables, it reads
\begin{equation}
d(t-s)= d_{\alpha\alpha}(t-s) = \frac{gN }{\tilde{T}} [ \langle \sigma_\alpha(t) \Phi_\alpha(s) \rangle - \langle \sigma_\alpha(s) \Phi_\alpha(t) \rangle ] \nonumber =0\   
\end{equation}
where $\sigma_{\alpha}= (1/N) \sum_i \sigma_i^{\alpha}$. We note that now there is no $v_0$ prefactor, nevertheless  when $v_0\to 0$ the deviation vanishes because the two correlation functions in the r.h.s. become identical due to the restoring of detailed balance. When considering the EPR, expressed in terms of the rescaled variables, when  $v_0\to 0$ it vanishes trivially because the time derivative of the adjacency matrix becomes zero (see Eq.~\eqref{EPR-full-parti}).